%% file: MainPaper.tex
\documentclass[Afour,sageh,times]{sagej}

\usepackage{url}

\usepackage[colorlinks,bookmarksopen,bookmarksnumbered,citecolor=red,urlcolor=red]{hyperref}
\usepackage{import}
\usepackage{makecell}
\usepackage{booktabs}
\usepackage{multirow}
\usepackage[table,xcdraw]{xcolor}

\usepackage[shortlabels]{enumitem}

\newcommand\BibTeX{{\rmfamily B\kern-.05em \textsc{i\kern-.025em b}\kern-.08em
T\kern-.1667em\lower.7ex\hbox{E}\kern-.125emX}}

\def\volumeyear{2016}

\begin{document}

% \runninghead{Anonymous Authors}
\runninghead{Ha, Kong, and Jhaver}

\title{Examining Racial Stereotypes in YouTube Autocomplete Suggestions}

% \author{Anonymous Authors}
% \author{Eunbin Ha\affilnum{1$^*$}, Haein Kong\affilnum{1$^*$}, and Shagun Jhaver\affilnum{1}}
% \affiliation{\affilnum{1}Rutgers University, New Brunswick, United States}
\author{Eunbin Ha\textsuperscript{1*}, Haein Kong\textsuperscript{1*}, and Shagun Jhaver\textsuperscript{1\dag}}
\affiliation{\textsuperscript{1}Rutgers University, New Brunswick, United States\\
\textsuperscript{*}Equal contribution.\\
\textsuperscript{\dag}Corresponding author.
}
\corrauth{Shagun Jhaver, Department of Library and Information Science, Rutgers University.}
\email{shagun.jhaver@rutgers.edu}

\begin{abstract}
Autocomplete is a popular search feature that predicts queries based on user input and guides users to a set of potentially relevant suggestions. In this study, 
we examine what YouTube autocompletes suggest to users seeking
information about race on the platform.
Specifically, we perform an algorithm output audit of autocomplete suggestions for input queries about four racial groups and examine the stereotypes they embody. Using critical discourse analysis, we identify five major sociocultural contexts in which racial information appears---\textit{Appearance, Ability, Culture, Social Equity, and Manner}. 
We found that the participatory nature of YouTube produces a multifaceted representation of race-related content in its search outputs, characterized by enduring historical biases, aggregated discrimination, and interracial tensions, while simultaneously depicting minority resistance and aspirations of a post-racial society.
% Our results show evidence of aggregated discrimination and interracial tensions in the autocompletes we collected and highlight their potential risks in othering racial minorities.
We call for innovations in content moderation policy design and enforcement to address existing racial harms in YouTube search outputs.
\end{abstract}


\maketitle
% \def\thefootnote{*}\footnotetext{These authors contributed equally to this work}\def\thefootnote{\arabic{footnote}}
% \corrauth{Shagun Jhaver, Rutgers University
% School of Communication \& Information}
% \email{sj917@rutgers.edu}

\renewcommand{\thefootnote}{\arabic{footnote}}
%%%%%%%%%%%  Introduction  %%%%%%%%%%%%%%

\section{Introduction}
\begin{figure}
    \centering
    \includegraphics[width=1\linewidth]{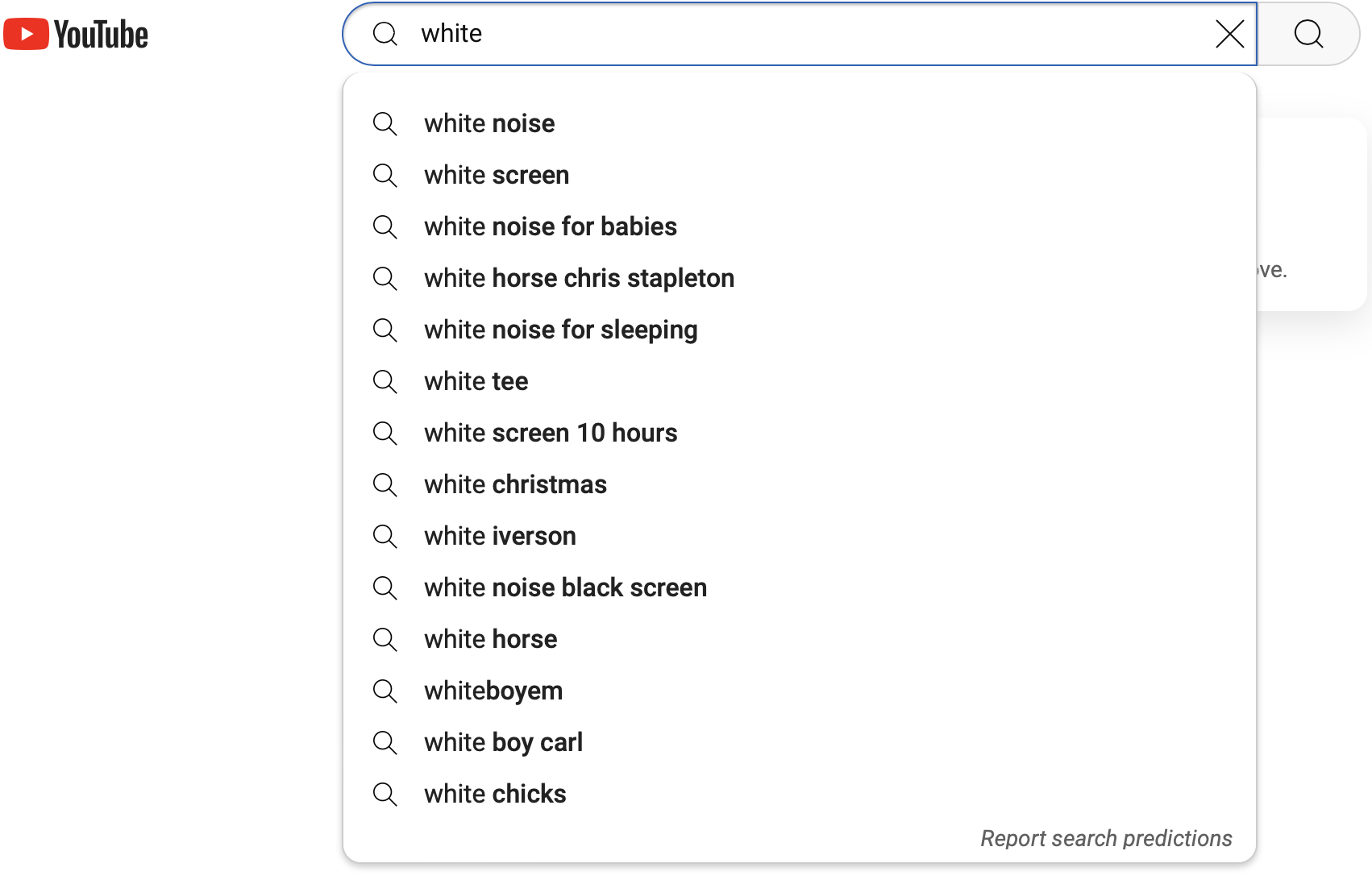}
    \caption{An example of YouTube's autocomplete function in action. Typing any keywords (e.g., "white") triggers a list of query options that users may select.}
    \label{fig:auto_ex}
\end{figure}

\textit{Autocomplete} is a technical feature that facilitates the online search process by using the first few entered keywords to predict and offer users a set of query suggestions~\citep{karapapa2015search} (Figure \ref{fig:auto_ex}). This feature, now ubiquitously available across all major search platforms, social media websites, and online marketplaces, is also available on YouTube,\footnote{\url{https://www.youtube.com/}} the world's largest video-sharing platform.
Autocomplete improves users' search experience by reducing typing by about 25 percent on average \citep{Sullivan2018}. 
However, the algorithmic `nudges' this feature offers have the potential to induce identity-based biases among YouTube searchers. This is because 
its autocomplete query suggestions 
%that constitute autocomplete results 
are shaped not only by the inputting user's previous YouTube activities and location but also by the wider popularity of such queries and the content uploaded by YouTube creators, who often tend to be wealthy and White~\citep{lennard2015youtube,wang2018game}. Thus, YouTube autocompletes could incorporate the prevalent biases and stereotypes that some people hold and expose them to other users \citep{lin2023trapped}. 
 
Such reproduction of stereotypes constitutes representational harm by narrowing demographic groups down to specific traits and exaggerating them \citep{leidinger2023stereotypes}. Negative stereotypes about oneself can influence one's well-being and productivity, whereas negative stereotypes about other groups can permeate public institutions and shape how society and policymakers treat those groups \citep{levy2009stereotype}. \citet[p.1020]{roy2020age} argue that the unsolicited suggestions offered by autocompletion induce ``incidental learning,'' a ``process of unconscious, unplanned absorption of contextual information,'' and are likely to influence users' short- and long-term beliefs and curiosity about different topics. Early investigations of Google autocompletion emphasized that autocompletion stereotypes, viewed and absorbed repeatedly, distort our worldview and perpetuate oppressive social relationships \citep{cadwalladr2016google,noble2018algorithms}. 

%\sandy{Sorry if this is a silly question. When I typed 'white' to YouTube prompt, I got different autocomplete results. Does everyone? If this is the case, how can we  generalize about autocomplete outputs wrt stereotyping?}
Most prior research on autocomplete suggestions centers on algorithmic outputs from general-purpose search engines, especially Google Search~\citep{al2020google, baker2013white, leidinger2023stereotypes, roy2020age}.  This research focus seems reasonable because Google Search has been the largest global search engine since the early 2000's.\footnote{\url{https://www.britannica.com/topic/Google-Inc}} However, the use of autocomplete to assist query inputs goes far beyond Google searching. For instance, YouTube, the platform we focus on in this study, has incorporated the autocomplete feature in its search function for many years. Indeed, if we count the number of searches made across all platforms, YouTube emerges as the second-largest search engine after Google \citep{brandwatch202057}, offering all its results in a video format. 
% According to official documentation, Google Search autocompletes are influenced by query language, location, trending interests, and user's past searches~\citep{google_autocomplete}, and YouTube autocompletes are based on user's search and viewing history on YouTube and trending searches in the area \citep{youtube_autocomplete}. 
While Google does not publicly describe the extent to which it uses the same data and algorithms to drive search autocomplete suggestions on Google and YouTube, official documentation suggests that YouTube autocomplete predictions are at least partly shaped by content specific to YouTube, including users' video watching history \citep{youtube_autocomplete}.
% This indicates that autocomplete results are shaped by user behaviors specific to each platform, suggesting that YouTube autocomplete reflects the characteristics of a video-sharing platform.
%\sandy{Would your readers assume that because YouTube is part of Google's parent company, both companies would have similar issues? Do they, in fact, use the same autocomplete algorithms? Perhaps clarify this here?}
However, YouTube autocomplete has attracted little scholarly attention, perhaps because of YouTube's public perception as a video-sharing platform rather than a search engine. 
%, particularly revolving around Google search feature
%is the largest video-sharing platform \citep{khan2022researching}

Unlike search engines, YouTube allows users to not just view, but also create content on its site~\citep{wotanis2014performing}.
Prior research on YouTube's participatory culture has characterized it as a critical form of expression, ``self-promotion, escapism, and utter play''~\citep{boxman2019practice}.
However, it is possible that YouTube's complex cultural fabric comprising both professional (i.e., traditional media) and amateur (i.e., independent individuals) content production~\citep{boxman2019practice} induces the types of biases that do not appear on search engines.
At the same time, independent YouTubers have the potential to create and share content that counteracts traditional biases and influence search outputs in positive ways.

YouTube algorithms construct and shape content creators' visibility on the site's search and front pages as well as in features like autoplaying of `related videos'~\citep{bishop2019managing}.
However, YouTube releases limited information about these algorithms to maintain a competitive advantage. 
This black-boxing of algorithmic governance makes accountability difficult~\citep{reynolds2024user}.
Thus, there are growing concerns about how cultural producers, who must navigate the increasingly algorithmic nature of content distribution on YouTube, risk losing their visibility and income~\citep{bishop2019managing}.
Given YouTube's unique participatory culture and the uneven power relationship between content creators or searchers and the platform, \citet{reynolds2024user} have called for examining YouTube algorithms' accountability in its specific platform environment.
Responding to this call, we focus on YouTube's search algorithms.

% Given YouTube's wide reach, the cultural influence of its offerings, and the unique content curation mechanisms that shape its search outputs (including both search results and autocomplete suggestions), we argue that it is crucial to 
% examine how YouTube's search algorithms might affect end-users' information gathering and belief formation processes. We begin to fill this vital research gap by 
Specifically, we explore what information YouTube users in the United States (U.S.) encounter through the search autocomplete suggestions. Since misinformation or limited information about racial groups, along with media stereotypes, can shape individuals' perceptions of race~\citep{covington2010crime}, \textit{we pose the following research question: 
what suggestions does the YouTube search feature provide to users exploring information about racial topics, and how may such suggestions shape users' understandings of race?} 
We focus on race here to extend prior conversations on the media's construction of race~\citep{chen1996feminization,squires2014post,walters2024memes}.

Importantly, the autocomplete feature not only provides information but may also contribute to the propagation of racial stereotypes.
%\sandy{Call it an 'algorithm output audit" to be more precise?}
\textit{We perform an algorithmic output audit of YouTube autocomplete suggestions regarding four racial groups---Whites, Blacks, Asians, and Hispanics---to surface such stereotypes.}
At the same time, we also attend to positive representations of racial minorities and attempts to counteract racial stigmas in these autocomplete suggestions. 

\section{Background and Related Work}
\subsection{Racial Stereotypes}

Since the 1950's, significant changes in America's political and social landscape have made racial discrimination illegal and rendered overt racial prejudice socially unacceptable. In response to these broad shifts, social scientists have examined how racial stereotypes have evolved in tandem. Recognizing these changes is crucial in intergroup relations research since shifts in stereotypes are often seen as a necessary prerequisite to reducing prejudice and fostering more positive intergroup interactions.

%Overall
% While the U.S. population is becoming increasingly more multiracial, mass media and political communications have sustained racial stereotypes.
% 
% but racial stereotypes and discrimination within the society continue to remain. 
% A majority of Americans believe that Black, Hispanic, and Asian populations in the U.S. face substantial discrimination based on their race, whereas White people face the least discrimination \citep{pew_discrimination}. According to the Pew Research Center, most Black Americans think that Black people are more likely to be described negatively in the news compared to other racial groups \citep{pew_black_news}. About 40\% of Latinos reported that they experience discrimination or unfair treatment in the U.S. because of their national/ethnic background \citep{pew_latinos_discrimination}. Additionally, Asian Americans experience the `forever foreigner' and `model minority' stereotypes, where they are often perceived as foreigners and successful minorities \citep{pew_forever_foreigner, pew_model_minority}. 
% 
% White 
Since Whites have long constituted the majority of the U.S. population, there exists a cultural tendency toward White ethnocentrism that contributes to shaping Whites positively and other racial groups negatively \citep{maykovich1972stereotypes}. 
% In reflection of this trend, the stereotypes about Whites predominantly tend to be tied to positive characteristics such as `intelligent' or `industrious' \citep{bayton1941racial, katz1933racial}. 
% Even, negative stereotypes (e.g., regarding natural physical ability) are based on the assumption that Whites' intelligence is superior to Blacks even if their physical skill is inferior to Blacks \citep{sailes1993investigation, stone1997white}. 
% At the same time, negative stereotypical attributes associated with Whites are less commonplace. 
% This is because, under the nation's salient racial hierarchy, the prevailing cultural norms and values of white Americans have been designated as `the standard' for a long time \citep{umana2020stereotypes}.
% In contrast, the term `Black' has been commonly used as a packaging label (e.g., Black crime) \citep{pickering2004racial}. 
Specifically, US racial stereotypes often manifest as placing Whites and Blacks at the top and bottom of the racial hierarchy, respectively, with other ethnic minorities,  such as Asians or Hispanics, in between these extremes \citep{song2004introduction}. 
% Thus, the American hierarchical social system contributes to the unidirectional flow of stereotypical attributions, leading to asymmetry in racial stereotypes.

% Black 
Historically, Blacks or African Americans have been negatively stereotyped with respect to a variety of sociocultural characteristics, including their personality, hygiene, and criminal conduct. Messages from mass media and political communications often reinforce these stereotypes~\citep{hurwitz1997public}.
Some scholars maintain that these distorted perceptions of Blacks are rooted in slavery \citep{lintner2004savage} and Christian mythology \citep{miller1995origins}. 
Whatever their origin, such racial stereotypes are dangerous because empowered non-minorities can exploit them to justify their privilege and continue to perpetuate harm and discrimination against 
`others.' For instance, racial segregation in the U.S. was historically based on biased perceptions about Blacks' personalities, intelligence, and appearance \citep{lintner2004savage}. 

% associated with many negative stereotypes in the U.S. Some scholars argue that these distorted perceptions of Blacks are rooted in slavery \citep{lintner2004savage} and Christian mythology \citep{miller1995origins}. Negative stereotypes for Blacks exist regarding a variety of sociocultural aspects, including their personality and criminal conduct. Messages from mass media and political communications often reinforce such stereotypes~\citep{hurwitz1997public}.
% % For example, Blacks were stereotyped as poor, lazy, aggressive, criminal, and unclean \citep{gilens1996race, lintner2004savage}. 
% Such racial stereotypes are dangerous in part because non-minority groups can exploit them to justify their privilege and perpetuate harm and discrimination against minority groups. For instance, racial segregation in the U.S. was historically based on biased perceptions toward Blacks' personalities, intelligence, and appearance \citep{lintner2004savage}. 
% On the other hand, positive stereotypes of Blacks' physical and athletic abilities \citep{simiyu2012challenges} compare them favorably to other racial groups.
% However, internalizing this stereotype can cause problems within the Black community, such as a lack of motivation for academic success among students \citep{czopp2010studying}.

% Asian 
Asians or Asian Americans, often described as a `model minority,' are stereotyped as being intelligent, hard-working, and academically and economically accomplished \citep{lee2008th, trytten2012asians}. While this model minority stereotype could be perceived as positive, \citet{lee2008th} argued that it can conceal the diversity among Asians, rouse interracial tensions, and discourage Asians from disclosing their problems and seeking assistance.  In addition to the model minority stereotype, Asians have a long history of being marginalized in the U.S. \citet{chen1996feminization} showed that the physical appearance of Asian male immigrants in the nineteenth century and the work they engaged in were often caricatured to build feminized and infantilized images; the mass media at that time reinforced this image by portraying them as wearing feminine clothes or engaging in housework  \citep{chen1996feminization}. Further, the U.S. mass media has often portrayed Asian women as exotic, submissive, and quiet~\citep{paner2018marginalization}. 

% Hispanic 
Another US minority, Hispanics, tend to be categorized as a single group despite their heterogeneous and complex ethnic roots~\citep{nelson2014structuring}. They were typically described in early Western films as highly emotional and violent individuals with untidy features, aka the `El Bandido' stereotype \citep{berg2002latino}. This stereotype has gradually evolved into the image of undocumented immigrants, contributing to racist discourses that portray them as criminals who are potentially threatening to American lives and jobs \citep{perez2015visualizing}. 

% How is your research adding to the insights described in this whole subsection?
Our work extends this prior understanding of racial stereotypes. We examine their \textit{current state} in the U.S. through the lens of how each racial group is described in YouTube autocomplete algorithm's suggestions. We aim to identify the perpetuating racial stereotypes reinforced by this culturally influential online platform and highlight recently emerging stereotypes.
In doing so, we also bring attention to the algorithmic biases.
%that have surfaced recently.

\subsection{Algorithmic Bias, Search Critiques, and Algorithm Audits}
% Algorithmic Bias - concept added 
As digital platforms grow, systems designers rely on algorithms to determine what information to present to whom. These algorithms and the interfaces created for them can signal the quality and relevance of different search results and shape people's perceptions of topics of societal importance~\citep{kay2015unequal}. However, both algorithms and interfaces can exhibit biases in their representation of information. 
While there is no universally agreed-upon definition of \textit{algorithmic bias}, researchers from various disciplines have addressed its different manifestations. In social science, scholars tend to focus on how algorithmic bias perpetuates social discrimination and affects social equity.

The research presented here can be placed within the broader critique of algorithmic biases in content moderation, though we focus on search feature outputs rather than AI-assisted regulation of potentially inappropriate social media posts~\citep{jhaver2019human}. Specifically, this article aligns with approaches that aim to assess the extent to which search outputs are: problematic; characterized by offensive, stereotypical, inappropriate, and discriminatory terms; and/or racist~\citep{rogers2023algorithmic}.
Unlike these approaches, we also document how the search outputs serve to dismantle racist stereotypes.
Our method of probing YouTube's search feature with specific queries could be positioned alongside approaches like 
%\sandy{Aren't you doing an algorithm audit? What does "positioned alongside" mean in this case?}
algorithm audits~\citep{sandvig2014auditing} and ethical hacking practices that search for vulnerabilities~\citep{roberts2019behind}.

Prior research has shown algorithmic biases in the operation of search engines regarding what they index, what they present to specific users, and---most relevant to our work---how their autocomplete suggestions reinforce social discrimination~\citep{kay2015unequal}.
For example,~\citet{baker2013white} showed that Google search produces a higher number of negatively stereotyped autocomplete suggestions for searchers with Muslim, Jewish, Gay, or Black identities.
\citet{lin2023trapped} collected autocomplete predictions from three leading search engines---Google, Bing, and Baidu---and showed disparities in their toxicity scores with respect to gender, race, and sexual orientation.
%\sandy{I'm not sure you ever address whether your findings were consistent with those for Google autocomplete research.}
~\citet{leidinger2023stereotypes} showed that autocompletes from Yahoo, another general-purpose search engine, include a large number of negative stereotypes for Latinos, e.g., portraying them as stupid and loud. 

Our work contributes to this line of research by 
% deploying a novel methodological approach and conceptual focus to understanding 
examining the prevalence of racial biases in YouTube search autocompletes. 
While prior studies have primarily focused on only the text of autocomplete suggestions~\citep{al2020google, baker2013white, leidinger2023stereotypes, roy2020age}, we analyze autocomplete texts in conjunction with the video results for each suggestion. This helps us present a more nuanced understanding of how YouTube users' interactions with its search feature may influence their perspectives. 
Our work is the first to deploy \textit{critical discourse analysis} (CDA) in addition to inductive analysis to uncover not only how language and discourse practices of autocompletes reflect problematic socio-cultural norms but also how they reinforce hegemonic whiteness and racist ideologies.

% on autocomplete algorithm audits and the measurement of algorithmic biases. 

It is possible that biased queries entered by users, in turn, could be influencing the autocomplete algorithm to produce stereotypical outputs.
However, given the black-box nature of autocomplete algorithm, the extent to which this happens remains unclear. 
Some recent evidence suggests that the impact of search volume on autocomplete suggestions may be lower than is generally assumed and, in some instances, prior user queries may not feature at all in certain suggestions~\citep{Graham2022Investigating}.
Thus, we largely focus in this article on the audit of autocomplete outputs rather than on inferring how biased user queries could influence the production of stereotypes in autocompletes.

\section{Methods}

\subsection{Development of Input Search Queries}

We began our inquiry by selecting the following four categories of racial groups: ``White,'' ``Black,'' ``Asian,'' and ``Hispanic.'' This selection was based on racial categories most frequently included in prior research on stereotypes or biases in search engine autocompletes and language models \citep{kirk2021bias, leidinger2023stereotypes}. Next, we developed a series of input queries based on combinations of group indicators and verb phrases. We leveraged the synthetic patterns of input queries employed in previous autocomplete studies \citep{al2020google, baker2013white, roy2020age} to guide this query creation. 
We combined the following verbs---\textit{is, are, does, do, can, should}---with racial group indicators in the form of declarative and interrogative sentences.
% To gain more explicit and comprehensive results related to each racial group, we excluded wh- question words in the input queries template of prior studies while instead combining the following specific verbs with racial group indicators in the forms of declarative and interrogative sentences: ``is, are, does, do, can, should.'' 
Table \ref{tab:query} shows the group terms and synthetic patterns we used to structure our input queries. In total, we curated 84 input queries for each of the four racial groups through this process.
% The total number of input queries is 84. We applied this set of input queries for each racial group, having 336 input queries in total.

\begin{table*}
  \centering
  \begin{tabular}{l p{6cm} | l}
    \toprule
         Category & Group terms & Synthetic patterns\\
    \midrule
        White & White/Whites, White person/people, White man/men, White boy/boys, White woman/women, White girl/girls & \multirow{0}{5.5cm}{Is a(an)/are [Group Terms] 
        \newline Does a(an)/Do [Group Terms] 
        \newline Can a(an)/Can [Group Terms]
        \newline Should a(an)/Should [Group Terms]
        \newline A(An) [Group Terms] is/are
        \newline A(An) [Group Terms] can/can
        \newline A(An) [Group Terms] should/should}
        \\
        Black & Black/Blacks, Black person/people, Black man/men, Black boy/boys, Black woman/women, Black girl/girls &  \\
        Asian & Asian/Asians, Asian person/people, Asian man/men, Asian boy/boys, Asian woman/women, Asian girl/girls & \\ 
        Hispanic & Hispanic/Hispanics, Hispanic person/people, Hispanic man/men, Hispanic boy/boys, Hispanic woman/women, Hispanic girl/girls &  \\ 
    \bottomrule
  \end{tabular}
        \caption{The structure of input queries used to guide our data collection.}
  \label{tab:query}
\end{table*}

\subsection{Data Collection}

\begin{figure}[h]
  \centering
  \includegraphics[width=1\linewidth]{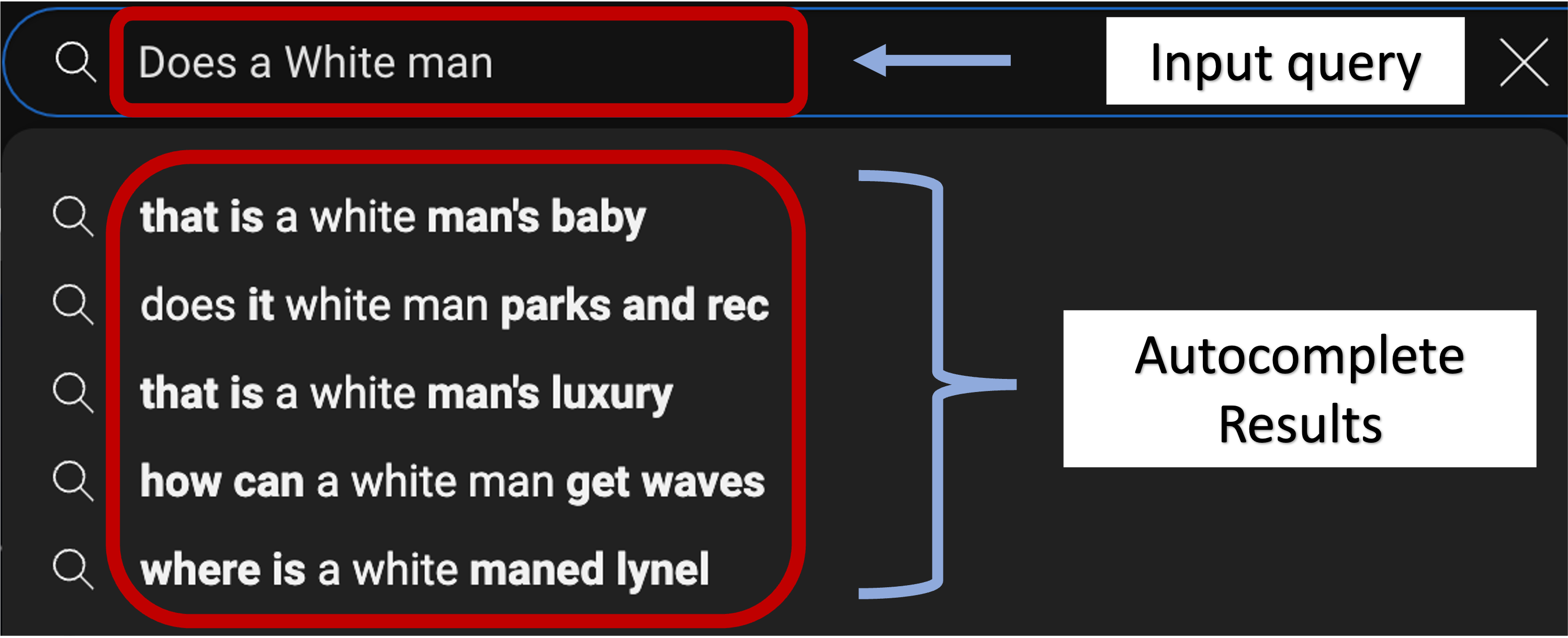}
  \caption{Screenshot of a YouTube search for one of our input queries. It shows five of the autocomplete results we collected that correspond to the query.}
  \label{fig:our_auto_ex}
\end{figure}

We collected data from July through August 2023 by gathering YouTube autocomplete suggestions for our input queries. Figure \ref{fig:our_auto_ex} shows an example of a YouTube search with a sample input query and its autocompletes; we collected such autocompletes for each query. 
We automated our data collection by developing Python scripts that simulated YouTube search querying actions. 
This automated process comprised three stages: 1) accessing the YouTube.com website, 2) entering an input query in the YouTube search box, and 3) collecting autocomplete results, if available, and storing them. We used Chrome as a web browser and accessed YouTube in incognito mode without logging in to any account to prevent the effects of personalization and browsing history. Our scripts opened and closed a new browser window for each query to ensure that the search history of other input queries did not influence the autocomplete suggestions for any single query.
  
% We automated our data collection by developing Python scripts that simulated YouTube search querying actions. 
%\sandy{Given that you sanitized your searching in this way, how representative is it of actual use?}

% VPN
We used a virtual private network to collect data from multiple locations in the U.S. to account for location-specific biases \citep{ribeiro2020auditing}. In particular, we collected YouTube autocompletes from five states: New Jersey, California, Georgia, Texas, and Washington. We conducted our data collection three times, each separated by a few days, to account for the temporal variations in algorithmic suggestions. Through these efforts, we attempted to build a geographically representative and temporally stable dataset of autocomplete suggestions across the U.S. 

Analyzing the data samples from three different waves and five different states, we found minimal differences among them. We integrated these samples and removed duplicate results to create a combined dataset for further inquiry. 
While this step precluded prioritizing more geographically and temporally stable autocompletes, it allowed us to work with a more comprehensive set of suggestions. Examining the resulting dataset, we also noticed that some autocomplete suggestions were not related to the racial groups. Thus, we removed the irrelevant autocomplete suggestions based on whether the autocomplete refers to non-human rather than human subjects. 
After removing duplicates, our
% Also, since our dataset still included the multiple identical autocompletes generated from different queries, we finally removed all the duplicate autocomplete. As a result, the 
dataset consisted of a total of 241 unique autocompletes. 

We describe more details of our data preparation process and descriptive statistics in the Supplemental Material.

\subsection{Data Analysis}
% % Quantitative analysis 
% We conducted a descriptive analysis to investigate the differences across racial groups in YouTube autocompletes. First, we compared the number of input queries that do not generate any autocompletes related to each racial group. Second, we compared the number of autocompletes of the racial groups. This analysis allows us to examine the differences in recommending autocompletion for racial groups on YouTube.

% content analysis 
We used inductive analysis \citep{elo2008qualitative} to examine our data and extract concepts and categories that describe the underlying phenomena.
% Two authors conducted qualitative coding aimed at inductively~\citep{elo2008qualitative} creating categories that span the four different racial groups. 
Two authors conducted a bottom-up open coding process on the platform `Dedoose' which allowed the authors to collaborate in real-time. 

We began by developing a situated understanding of each autocomplete suggestion by examining about ten videos displayed as top search results when selecting that suggestion. 
This examination helped us recognize the contextual and cultural meaning of autocomplete suggestions, and we leveraged this nuanced understanding to develop our open codes. 
Since we were primarily interested in the autocomplete texts, we did not create open codes to describe the video content; instead, we created codes and memos that delineated the racial biases, tensions, and struggles suggested by the corresponding autocompletes and the sociocultural contexts in which they occurred.
This careful, context-driven coding process was important because some of the autocomplete results were closely related to memes or video content in their corresponding search results. 
For instance, the autocomplete “when an asian is kidnapped” was related to comedy content created by an Asian YouTuber. Thus, it was important to search for and understand the situated meanings of autocompletes rather than interpret them in isolation. 
However, in many cases, the resulting videos did not directly relate to or provide any additional context for the autocomplete text; we suspect that in such cases, the autocomplete texts could have been influenced by the search patterns of YouTube users rather than the content of YouTube videos. 

Then, we developed the common themes arising from our open codes and their corresponding autocomplete data by grouping similar codes into conceptual categories. At this point in the process, when any pre-existing categories seemed not to directly align with an emergent theme, we carefully considered whether to introduce a new category, revise an existing category, and/or create a subcategory. 
This process was iterative and we compared our emerging categories with one another and with the underlying data at each stage of our analysis. 
Some autocompletes represented miscellaneous contexts that did not warrant inclusion in or creation of a separate category; we excluded such autocompletes from further analysis. In total, we grouped 217 autocompletes into a two-tier category structure, where the main categories represented the overarching sociocultural contexts in which racial biases manifest, and the subcategories represented more specific discursive events within these contexts. The five main categories that emerged through this process included \textit{Appearance, Ability, Culture, Social Equity, and Manner} (see Table~\ref{tab:results}).

Throughout the process of qualitative coding and categorization of autocomplete results, we incorporated \textit{critical discourse analysis} (CDA). CDA is a systematic approach to examining a text, event, or discursive practice to identify how it originates from and is affected by power relations \citep{fairclough1993critical}. We employed this framework to analyze each autocomplete as concurrently a text (language produced in a discursive event), a discourse practice (the production, distribution, and consumption of a text), and a sociocultural practice (hegemonic struggles that ideologically shape a discursive practice) \citep{fairclough1995critical}. 
This meant that during data analysis, our focus was less on examining the lexical meanings of each autocomplete suggestion than on the hegemonic function the suggestion performed or indicated within the sociocultural practices represented by our five main categories.
For many autocompletes, our attention to their corresponding video results helped us characterize their text production, consumption, and underlying hegemonic struggles.
Using this approach, we aimed to identify various facets of each discursive event, thereby revealing particular power dynamics reflected in that event. We engaged in regular discussions about how the autocomplete suggestions in our data reproduce and legitimize social inequalities, and our coding sought to identify various forms of domination, biases, and resistance.

%%%%%%%%%%%  Results  %%%%%%%%%%%%%%

\section{Results}
Our analysis generated five main categories representing distinct contexts in which we observed racial biases: \textit{Appearance, Ability, Culture, Social Equity, and Manner}. Under these five main categories, we identified a total of 13 subcategories (Table \ref{tab:results}).
%\sandy{Please adjust cell size of rows to avoid wrapping in central row.}
\begin{table*}[htpp]
  \centering
  \begin{tabular}{l l l r}
    \toprule
    Main category & \ Subcategory & \ Description & \% (n)\\
    \midrule
        Appearance 
             % & Facial Hair & The hair on the face %(i.e., beard)  \\
             & Personal Hygiene 
             & The activity of cleaning the body and maintaining a good appearance  & 3 (6)\\
             & Skin Tone 
             & The skin color originating from racially different genetic factors & 5 (10)\\
        Ability 
            & Talent
            & One's abilities for artistic or physical performances & 22 (47)\\
            & Financial & One's abilities to purchase goods or services & 2 (4)\\
            & Intellectual 
            & One's abilities to learn and study & 2 (4)\\
        Culture 
            & Cultural Heritage 
            & Conventional cultural elements of one society with a shared identity & 21 (45)\\
        % & & (i.e., cultural practices and cultural attires) \\
        & Ethnic Humor 
            & The humor based on racial, ethnic, or national stereotypes & 13 (28) \\
        % & & within a specific region or cultural community  \\ 
        & Language 
            & How someone communicates verbally & 6 (12)
        \\
        % & & (i.e., one’s voice or accent) \\ 
        % & Lifestyle & The typical pattern of individuals' or groups' lives \\
        % & & (i.e., daily behavior or habits)  \\
        & Relationships 
            &  How people interact and connect with each other & 9 (19) \\
        % & & (i.e., romantic relationship) \\
        Social Equity 
            & Diversity/Inclusion 
            &  Practices of embracing socially diverse groups of people & 2 (5) \\
            & Racial Justice 
            &  Pursuit of fighting against injustice toward a certain racial identity & 10 (22) \\
        Manner 
            & Aggression 
            & Antisocial and violent traits & 5 (11) \\
            & Inappropriate Behavior & Being idle or behaving in a socially inept way & 2 (4)\\
    \bottomrule
  \end{tabular}
        \caption{Categories of racial stereotypes that emerged in YouTube autocomplete suggestions. \% and n refer to the percentage and number of autocompletes belonging to each subcategory in the analyzed data, respectively.}
        \label{tab:results}
\end{table*}

\subsection{Appearance}

% \textbf{\textit{Facial Hair}}
% We found autocompletes asking about the possibility of growing a beard for Asian men, ``Can Asian men grow beard?''. This autocomplete appeared only for Asian men. Given that a beard is related to masculinity while the clean-shaved face reflects sexual immaturity, this result contains the feminized stereotype of Asian men as less masculine \citep{chen1996feminization}. 

\begin{figure}
    \centering
    \includegraphics[width=1\linewidth]{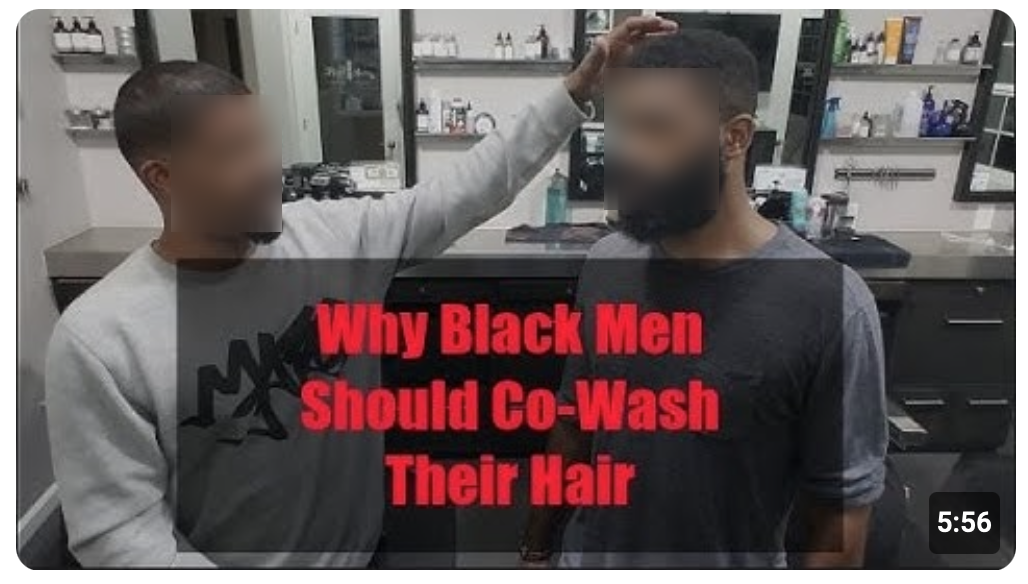}
    \caption{A YouTube video suggested by the autocomplete “how often should a black man wash his hair.”}
    \label{fig:hygiene-2}
\end{figure}

\textbf{\textit{Personal Hygiene.}}
This category is notable because only Blacks were subject to autocompletes related to personal hygiene. These autocompletes referenced Black people washing their hair or beard (``how often should a black man wash his hair,'' ``how often should a black woman wash her hair,'' ``how often should a black man wash his beard'') and shaving (``should a black man shave everyday''). The resulting videos featured creators presenting their shower or shaving routines, e.g., Figure \ref{fig:hygiene-2} shows the preview of a video resulting from the suggested autocompletes, which contains advice for preventing white flakes in hair.\footnote{\url{https://www.youtube.com/watch?v=CZCuI4x1sUc}} 
Another autocomplete asked, ``do black people get lice.'' 
It is possible that these autocompletes reflect highly sought-after searches by Black people for the information needed to take care of textured hair.
However, these questions are raised only for the Black group, raising the possibility that they reflect the racial stereotypes of Black people having poor hygiene. Historically, a deep-rooted racial stereotype maintained that dark or non-white skin was associated with uncleanness \citep{brown2009foul}; indeed, this stereotype served as a rationale for racial segregation in the U.S. \citep{lintner2004savage}. These YouTube autocompletes suggest that the negative stereotypes about Blacks' hygiene still persist. 

\textbf{\textit{Skin Tone.}}
Similar to `Personal Hygiene,' Blacks were the main subjects of autocompletes in the `Skin Tone' category. Autocompletes related to skin tones included ``can a black person become fair.'' These bleaching-related autocompletes reflect both an interest in lighter skin and the implicit negative view of darker skin \citep{hunter2011buying}. We observed no autocompletes related to the action of changing natural skin tones for other races. Other autocompletes show how Black skin is perceived as unique and distinct from other races. Examples include autocompletes asking about the need for Black people's skin protection (``should black people wear sunscreen,'' ``do black people get sunburn'') and changes in their natural skin tones (``can black people blush,'' ``can black people tan''). 
% Only one autocomplete for other groups was shown (``That is a White man's baby''). Given that the video content corresponding to this suggestion was linked to the Black's dialogue ``That is a Slavic baby, a Viking from Iceland'' toward a light-skinned baby from a movie called ``The School Dance,'' the autocomplete reflects the context of doubting the child's biological parents based on physical appearance. 
These results suggest both an otherization and a general lack of awareness about Black skin characteristics.  
However, some videos for these autocompletes showed creators dismantling stereotypes about Black skin, e.g, in one such video,\footnote{\url{https://www.youtube.com/shorts/tMkpBnmNy7o}} a Black creator applied blush to her skin.

%\sandy{I'm not sure 'ignorance' is the right word; could simply be 'uninformed.' }

\subsection{Ability}

\textbf{\textit{Talent.}}
Our analysis revealed racial stereotypes associated with talent in music, dance, and sports that mainly addressed White and Black groups. For White groups, the autocompletes related to their artistic talent included both positive (``white girl can rap,'' ``white girl can sing") and negative (``white people can't dance'') elements. Regarding physical talent, several autocompletes reflect stereotyped perceptions that Whites are not as athletic as Blacks (``white men can't jump,'' ``white boys can't dance''). Indeed, the videos that such autocompletes guided users to related to a 1992 sports comedy movie called ``White Men Can't Jump," which leverages the widely held stereotype that Black race is a positive indicator of physical ability \citep{felson1981self}. Autocomplete suggestions linked to this movie's title decades after its release date suggest a cultural appetite to mock or stereotype White athletes as being inferior in their physical performance.

%both positive and negative
Similarly, most autocompletes for Blacks referenced their artistic or physical ability in both positive (``black people can dance to anything,'' ``black man can sing'') and negative (``black man can't dance,'' ``black man can't play basketball'') terms. Notably, we found some autocompletes related to Blacks' exceptional athletic abilities (``are black people more athletic''), which aligns with the previously reported stereotype~\citep{biernat1994shifting} that Blacks are athletically superior to Whites. 
Those who did not meet such expectations were negatively stereotyped. For example, the autocompletes ``black man can't dance,'' or ``black man can't play basketball'' reflect a stereotype that Black people should be good at dancing or basketball naturally, and the resulting videos mock Blacks who are not good dancers or sportsmen. In sum, these findings show that YouTube autocompletes reflect racial stereotypes based on widely held beliefs about the relationships between race and physical abilities.

\textbf{\textit{Financial.}}
Only Blacks were subjects of autocompletes that belonged to financial ability. All autocompletes here were associated with a lack of financial resources and acumen, which reinforces poor Blacks as the representative public image of poverty. Interestingly, this is consistent with how network TV news and weekly news magazines have historically grossly overrepresented Blacks in their portrayal of poor people as a whole~\citep{gilens1996race}. Autocompletes in this category reflected the stereotype that Blacks cannot purchase or own expensive products, such as fancy cars or clothes (``black man cannot own g wagon,'' ``black woman can't buy a dress,''  ``a black man can't have a suitcase''). 

\begin{figure}
    \centering
    \includegraphics[width=1\linewidth]{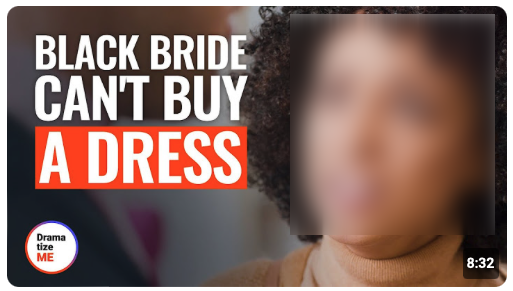}
    \caption{A YouTube video suggested by the autocomplete ``black woman can't buy a dress.''}
    \label{fig:dress}
\end{figure}

While these results superficially reinforce the long-standing stereotypes of Blacks as poor,
% the context of the autocompletes and the resulting video often counter this stereotype. FTwo autocompletes are a line from a YouTube video and a stand-up comedy show that criticizes the views that Blacks are poor. Also,  
some video results for these autocompletes counter this stereotype. 
For example, 
the top video result for ``black woman can't buy a dress'' portrays a White shop assistant and manager treating a Black woman attempting to buy a bridal gown at a shop differently than they treat a White woman buying a gown (Figure \ref{fig:dress}).\footnote{\url{https://www.youtube.com/watch?v=XMcRS2-la3E}} 
%\sandy{I watched this video to figure out what was being written here since it was unclear to me. The stereotype being addressed was the lack of black hygiene, i.e., they didn't want the Black woman to try on the dress, although it was originally presented as an ability to pay issue. Yes, the woman could afford the dress, but that wasn't the point.}
This video undermines the stereotypical attitudes toward Black people regarding their financial ability (it also critiques the assumptions of criminality and poor hygiene associated with Blacks). Similarly, resulting videos for other autocompletes about financial ability dismiss the assumption that Black people are poor by explicitly showing that Black people can and do own expensive products. 
%\sandy{I'm not sure what the following means. I suggest that the entire paragraph be rewritten for clarity.}
This shows that in some cases, users who review and watch the search results may get exposed to video content that helps diminish negative stereotypes of Blacks regarding financial ability. 

\textbf{\textit{Intellectual.}}
Intelligence-related autocompletes appeared for Asians and Blacks. In the U.S., Asians have historically been perceived as intelligent, hard-working, and pursuing high educational achievements \citep{trytten2012asians}. 
Autocompletes such as ``are asians smarter'' reflect this model minority stereotype. 
Unlike Asians, Blacks have often been negatively stereotyped regarding their intellectual and academic abilities~\citep{lintner2004savage,miller1995origins}. 
In line with this, we found autocompletes regarding Blacks' intelligence: ``black people can't name one african country,'' and ``black man can't say beginning.'' Videos for the first autocomplete show the lack of knowledge of some Black people about Africa, which suggests their ignorance of Blacks' historical or ethnic roots. The second leads to a video of a South African President, who is Black, repeatedly mispronouncing his words when attempting to say ``in the beginning.'' These cases show how videos uploaded to YouTube position individual mishaps to denigrate racial groups under the guise of humor. YouTube's automated use of such content to build autocomplete suggestions further removes necessary context and can thus perpetuate racial stereotypes.

\subsection{Culture} 

\textbf{\textit{Cultural Heritage.}}
Our analysis reveals that YouTube autocompletes reflect long-standing debates about \textit{cultural appropriation}, described as ``the use of one culture’s symbols, artifacts, genres, rituals, or technologies by members of another culture'' \citep[p.476]{rogers2006cultural}. We found that concerns about such appropriation appear in YouTube autocompletes related to several aspects of cultural heritage, such as cultural practices and culturally specific attire.

First, we found autocompletes that indicated appropriation of cultural practices, highlighting asymmetric power relations between racial groups. For example, several autocompletes related to White groups coopted 
%revealed an implicit act of consumption 
indigenous Black cultures while belittling its native contexts (e.g., ``white girl can dance african''). Other autocompletes revealed how  White people could appropriate specific aspects of popular Asian culture, e.g., ``can a white guy won [sic] an asian beauty pageant,'' ``can a white person become a kpop idol.'' The keyword `Black' elicited one autocomplete related to the adoption of Asian cultural elements (``can a black person become a kpop idol'').

\begin{figure}[h]
    \centering
    \includegraphics[width=1\linewidth]{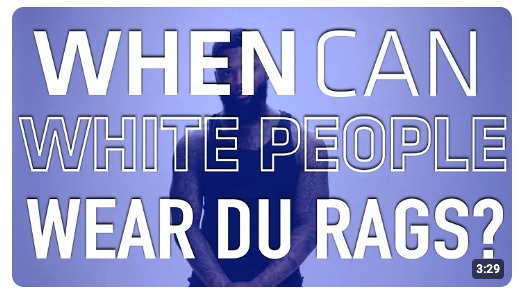}
    \caption{A YouTube video suggested by the autocomplete ``can a white person wear a durag.''}
    \label{fig:Wdurag}
\end{figure}

Second, autocompletes associated with culturally specific attire and beauty standards also indicated instances of cultural appropriation, where members of the majority group coopted traditional garments, hairstyles, or makeup practices of other minority groups. For example, autocompletes related to garments or hairstyles for White groups included traditional styles with deep origins in Asian or Black culture, e.g., ``can a white person wear a kimono,'' ``can a white person wear a durag,'' ``can white girls get braids,'' and ``can a white boy get waves'' (Figure \ref{fig:Wdurag}).\footnote{\url{https://www.youtube.com/watch?v=gSZ8U64EL-Q}}
%In contrast, autocompletes for the Asian or Black group did not produce any relevant results mentioning their own traditional garments or hairstyles. 
% Textual autocomplete suggestions related to  culturally specific attire guided users to a list of videos with voices and perspectives of diverse racial groups on cultural appropriation.
Additionally, we found instances of \textit{blackfishing}, i.e., the practices of modifying one's appearances physically or digitally or using terms and language patterns linked to Ebonics for racial commodification~\citep[p.1]{stevens2021blackfishing}.
% ; this phenomenon is particularly driven by the acts of White women mimicking Black women's looks to achieve economic gains \citep{cherid2021ain}. 
For example, we found autocompletes for White groups that included references to White people attempting to look like Blacks (e.g., ``the white woman who turned black''). Some videos linked by these autocompletes displayed the television coverage of Martina Big, a German model who underwent a perma-tanning procedure to darken her skin color.

In contrast, for Asians and Blacks, autocompletes related to cultural attire rarely reflected acts of cultural appropriation toward other cultures. Instead, the autocompletes for Blacks were related to changing their natural hairstyle (``how can a black person get curly hair,'' ``can a black man straighten his hair'') or achieving a popular hairstyle or makeup (``how to do a black person's hair,'' ``how to do a black person's makeup,'' ``how to do a black man bun''). 

The autocompletes for Asians often included the term ``Asian Baby Girl (ABG)'' (``what is an asian baby girl,'' ``what is an abg asian baby''). \textit{Asian Baby Girl} is a slang term referring to ``a young Asian woman who displays stereotypical traits,  such as enjoying clubbing, wearing excessive makeup and tattoos, drinking bubble tea, wearing revealing clothes, etc'' \citep{wiktionary-abg}. Though the popularity of such language can subvert stereotypical images of Asian women as submissive or quiet, it also associates them with negative attributes, like aggression and violence \citep{wu2023study}. Thus, autocompletes about ABG could increase awareness about the diversity of Asian women but, in doing so, risk generating new negative stereotypes about them.

In sum, our analysis reveals that YouTube autocomplete suggestions tend to be unidirectional, i.e., supporting ways that dominant racial groups (usually White) appropriate the cultural heritage of racial minorities but not vice versa.

\begin{figure}[ht]
    \centering
    \includegraphics[width=1\linewidth]{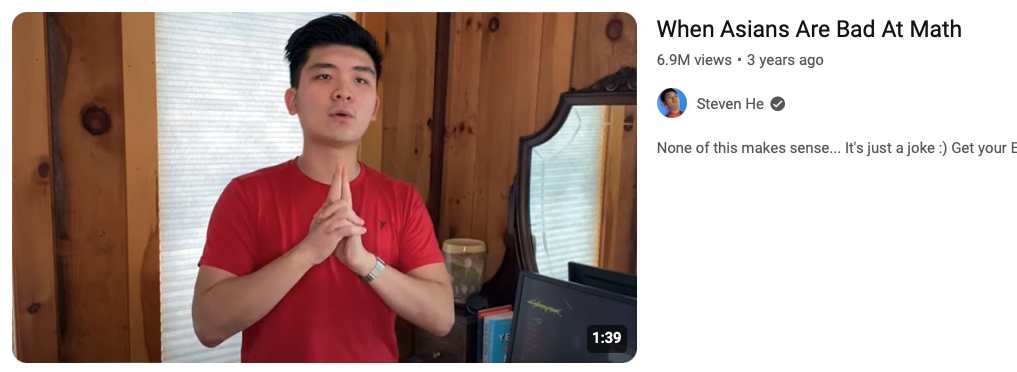}
    \caption{A YouTube video suggested by the autocomplete ``when an asian is bad at maths.''}
    \label{fig:steven}
\end{figure}

\textbf{\textit{Ethnic Humor.}}
Schutz defined ethnic humor as ``humor directed at racial and nationality groups, denigrating alleged attributes of those groups'' \citep[p.167]{schutz1989sociability}. Ethnic stereotypes or shared beliefs toward specific ethnic groups are the basis of such humor \citep{boxman2015ethnic}. In our analysis, Asians have the largest number of autocompletes in this category. Most of these autocompletes led to content created by the Chinese-Irish comedian Steven He O'Byrne (12.3 million subscribers).\footnote{\url{https://www.youtube.com/@StevenHe}} O'Byrne's videos usually parody Asian dads and mock East-Asian parenting stereotypes. For instance, the video content corresponding to the autocomplete ``when an asian is bad at maths'' shows O'Byrne portraying Asian parents who have a high expectation for their children's academic achievements and are harsh toward them when grades do not meet their standards (Figure \ref{fig:steven}).\footnote{\url{https://www.youtube.com/watch?v=6lMpjmPiTWE}} Other autocompletes, like ``when an asian is kidnapped,'' ``when an asian is president,'' and ``when an asian is your substitute,'' lead to videos with similar stereotypical images of Asians. While these videos presumably aim to generate clicks and their associated revenue through humor, they %may contribute to perpetuating traditional 
reproduce stereotypes by portraying Asians as alien or different and endorsing the `model minority' myth, which characterizes Asian Americans as being financially and educationally more advanced than other minority groups \citep{lee2008th}.

% \textbf{\textit{Lifestyle}}
% Autocompletes in this category revealed racially stereotyped attributions to lifestyle choices. The U.S. media and popular culture often portray White girls as frivolous, characterizing them as `blonde, text speak, creaky voice, selfies, or Starbucks' \citep{slobe2018style}. In line with this, our analysis found that YouTube autocompletes regarding White girls often appeared alongside branded products, such as Starbucks (“A White girl goes to Starbucks”). This shows how YouTube autocompletes may reflect preexisting racial stereotypes that position White girl personas as symbolic of material consumerism \citep{slobe2018style}. Meanwhile, autocompletes for Asian females were often associated with curiosity about the lifestyles of Asian women (“What does an Asian woman do when she comes home from work). These autocompletes guided users to a list of videos that show the daily routines of a diligent and self-reliant female. Again, such content reflects stereotypical expectations based on the model minority myth of Asian Americans \citep{lee2008th}.  

\textbf{\textit{Language.}}
Language attributes (such as accents) tend to be understood as a strong indicator of ethnic categorization \citep{rakic2011blinded}. We found several autocompletes related to accents or speaking styles of each racial group. For example, the autocompletes for speaking styles included, ``how to do a white girl voice,'' ``how to do a white person voice,'' and ``how to do a black man voice.'' Autocompletes for accents referenced the unique mode of pronunciation of certain racial groups (``black man can't pronounce,'' ``what does an asian accent sound like,'' ``all latino accents,'' ``woman with spanish accent''). Videos resulting from such autocompletes often depicted individuals mimicking or mocking the voices and accents of other racial groups. This suggests that the multicultural aspects of U.S. languages are still being leveraged to otherize specific groups. 

\textbf{\textit{Relationships.}}
Many autocompletes in this category highlight anxieties about interracial relationships. For example, some autocompletes reflect curiosity or potential worries about relationships between White women and people with darker skin (``do white girls like brown guys,'' ``do white girls like latinos''). Similarly, many autocompletes referenced relationships between Black people and women from different cultures (``can a black man marry an arab woman''). These results often portray interracial relations as incongruous and reflect an unfavorable view of darker skin in the relationship contexts. 

The most severe tension was found regarding the interracial relationships or marriages between White and Black individuals (``can't date a black woman until my grandpa died,'' ``black people can't marry meme,” ``black people can't marry white people''). These results indicate that cultural remnants of the historical prohibitions of interracial marriage and fears around miscegenation persist \citep{pascoe1991race}. There was only one result for Asians in this category: ``asian men are bitter.'' Though seemingly unrelated to romantic relationships, the top resulting videos of this autocomplete had titles such as ``Asian men are least desired.'' These videos depicted the struggles of Asian men with dating and pursuing romantic relationships, thereby perpetuating their historic feminization and infantilization~\citep{chen1996feminization}.

\subsection{Social Equity}

\begin{figure}[ht]
    \centering
    \includegraphics[width=1\linewidth]{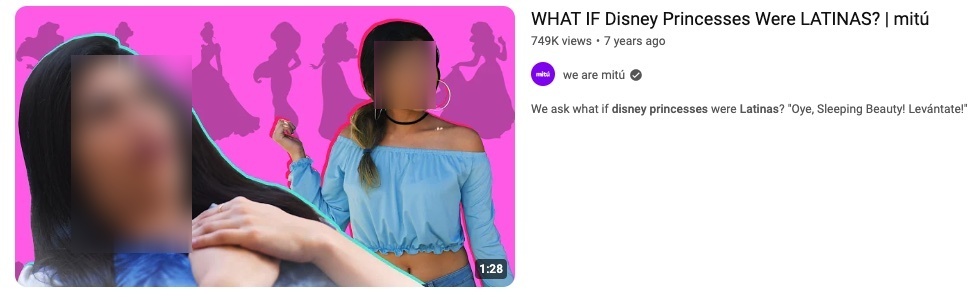}
    \caption{A YouTube video suggested by the autocomplete ``is there a hispanic disney princess.''}
    \label{fig:Hispanic3}
\end{figure}

\textbf{\textit{Diversity/Inclusion.}}
Whites, Blacks, and Hispanics had autocompletes associated with the issues of diversity and inclusion. Curiously, most of these autocompletes referred to the underrepresentation of racial minorities in popular culture (``can a black man be captain america,'' ``is there a hispanic disney princess''; see Figure \ref{fig:Hispanic3}).\footnote{\url{https://www.youtube.com/watch?v=AQLUeshn3nE}} This reflects a cultural appetite for seeing racial diversity and encouraging inclusive casting decisions in mainstream media. While whitewashing has historically been salient in the entertainment industry, the percentage of White characters depicted on-screen has slightly declined in recent years, while the depiction of Asian characters has significantly increased \citep{smith2023inequality}. However, no notable changes have occurred in the prevalence of Black or Hispanic characters over the same period \citep{smith2023inequality}. Seemingly in response to these trends, YouTube autocompletes in this category reveal frustrations about persistent inequalities in the racial representation of Black and Hispanic character portrayals.

\begin{figure}
    \centering
    \includegraphics[width=1\linewidth]{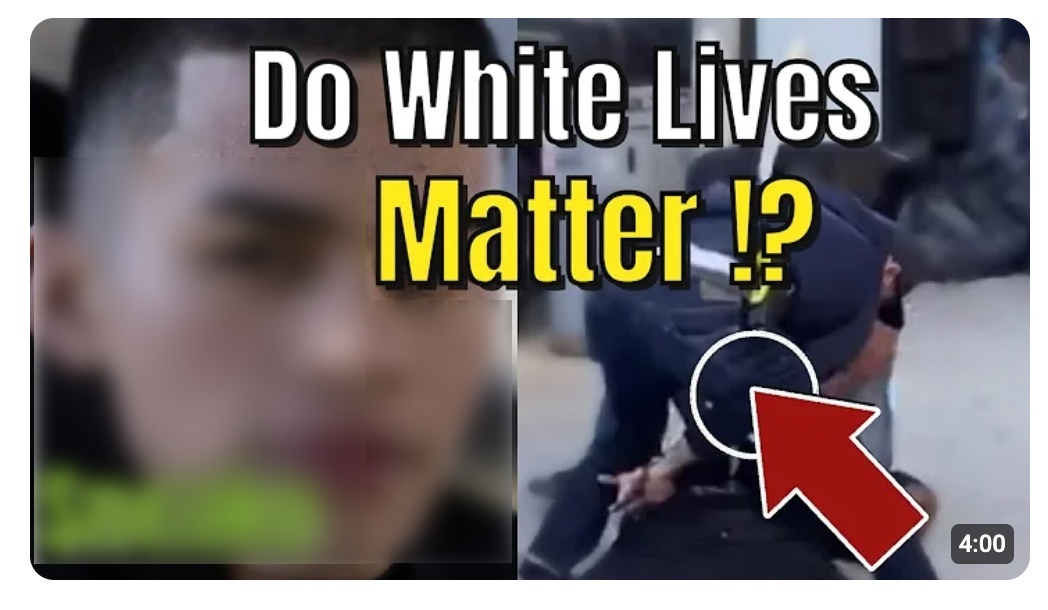}
    \caption{A YouTube video suggested by the autocomplete ``do white lives matter one minute.''}
    \label{fig:whitematter3}
\end{figure}

\textbf{\textit{Racial Justice.}}
We observed autocompletes in this category only for the White and Black racial groups. First, we found autocompletes that referenced `White Lives Matter,' a racist counter-movement that emerged in response to the `Black Lives Matter' movement (``do white lives matter one minute,'' ``do white lives matter shorts'') (Figure \ref{fig:whitematter3}).\footnote{\url{https://www.youtube.com/watch?v=GlMYUFDb8qg}} `White Lives Matter' 
%represents an active resistance to the `Black Lives Matter' movement, as it 
uses a variation of the Black Lives Matter core slogan to weaken the latter's message about addressing unwarranted police violence against Blacks \citep{goodman2023alternative}. 
% \textbf{`White Lives Matter' supporters have also been criticized by the Anti-Defamation League for its associations with racism, and, in 2019, it was designated as a hate group by the Southern Poverty Law Center.} Despite this controversy, 
The slogan `White Lives Matter' appeared verbatim in YouTube autocompletes. Despite video results for the query `White Lives Matter' displaying diverse voices and criticisms on contested topics, 
%\sandy{As a researcher, try to avoid injecting your feelings or emotions into your writing (e.g., "somewhat reassuringly" that preceded the first clause.}
%However, it is concerning that the 
text suggestions of YouTube autocomplete itself were solely generated based on the messages of `White Lives Matter,' and no analogous autocompletes were associated with `Black Lives Matter' messaging. This indicates that the autocompletes curated by YouTube do not always balance perspectives from opposing sides of an issue and, in this case, give short shrift to the side combatting racial disparities.

One autocomplete indirectly reflected the perspective of Blacks on racial inequality: ``can't keep a black man down,'' a lyric excerpt from 2Pac's song ``Trapped,'' which addresses police brutality and harassment faced by African Americans \citep{gaines2022ain}. This autocomplete and its resulting videos highlighted ongoing racial inequality and police brutality toward African Americans in the U.S.

\subsection{Manner}

\textbf{\textit{Aggression.}}
Our results reference perceptions of Blacks and Asians as being aggressive. Blacks had the highest number of autocompletes, which often described them as dangerous (``pretend there is a black man chasing you,'' ``run like a black man is chasing you'') or criminal (``a black woman is accused of shoplifting''). This aligns with long-held negative stereotypes of Black violence and aggression~\citep{gilens1996race}. We also found one autocomplete that suggests a desire to circumvent this stereotype: ``every black man should have a latte.'' While this does not relate to aggressive traits at face value, the top resulting video shows a Black stand-up comedian poking fun at the violence stereotype by recommending Black men hold a latte to appear `safe' to law enforcement officials.  Additionally, we found aggression-related autocompletes for Asians: ``asian man goes crazy.'' The top resulting video for this shows an Asian male game streamer screaming violently while playing games. Other videos show Asian men engaging in gun violence. While such aggressive images are typically not associated with Asian men in the U.S., these videos depict Asian men as not always docile and potentially violent.

\begin{figure}
    \centering
    \includegraphics[width=1\linewidth]{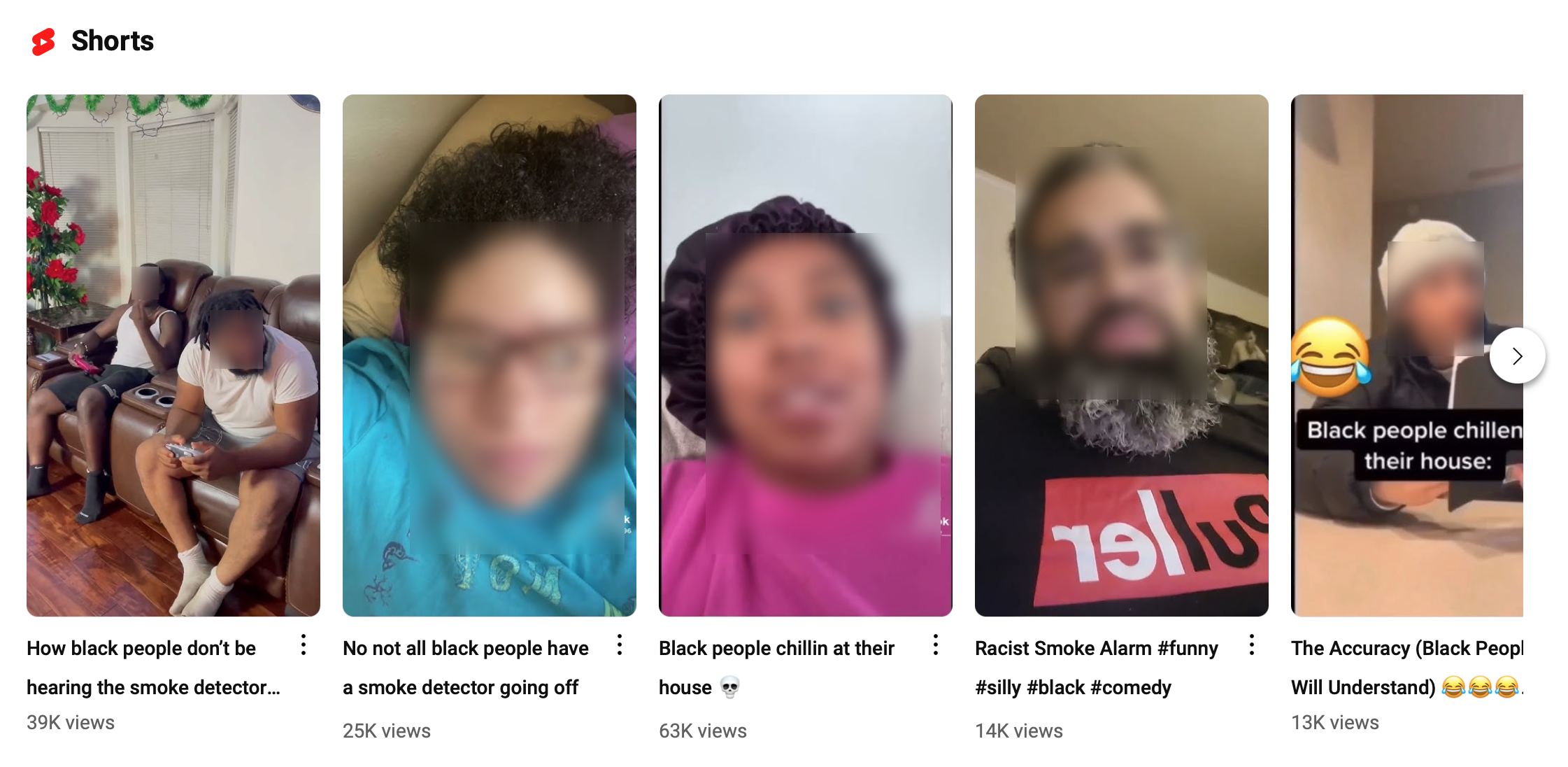}
    \caption{A series of YouTube shorts suggested by the autocomplete ``black people can't hear smoke detectors.''}
    \label{fig:smoke}
\end{figure}

\textbf{\textit{Inappropriate Behavior.}}
This category contains autocompletes that describe Black people as lazy or behaving inappropriately in social settings. One longstanding negative stereotype of Black people maintains that they are lazy, which is also posed as an explanation for other undesirable attributes, such as their disproportionate poverty \citep{gilens1996race}. The autocomplete ``black people can't hear smoke detectors'' depicts this stereotype; it led to videos that show Black people as being too lazy to change the low battery of smoke detectors even though their alarms kept sounding. Figure \ref{fig:smoke} presents a series of YouTube shorts that show how creators promote this stereotype via memes.

Several autocompletes describe Black people as lacking the self-control needed to behave appropriately in social settings. For instance, this category included ``black man cant [sic] hold his laugh,'' and ``black man can't stop laughing.'' The resulting videos show instances where a Black man cannot refrain from laughing in a serious situation where laughter is clearly inappropriate. 

% \newline \\

%%%%%%%%%%%  Discussion  %%%%%%%%%%%%%%

\section{Discussion}
We began this study with the goal of examining the information that YouTube's search feature provides to users seeking video content on race-related topics. 
As expected, we found the presence of racial biases and stereotypes in search autocomplete outputs, revealing the critical problems with search moderation. However, we also found that the participatory nature of YouTube has contributed to a more complex picture of race-based information in its search outputs, marked by desires for a race-neutral society, condemnations of historical oppressions, and curiosities about other races. We discuss these complexities below.

First, our analysis reveals that race-related autocompletes embody troubling stereotypes regarding Blacks' personal hygiene, financial abilities, language use, aggression, and laziness. 
% These stereotypes echo enduring anti-Black tropes rooted in colonialism, slavery, and segregation. 
Many search outputs reinforce historical biases against Blacks (e.g., the poor Black, the ignorant Black, the criminal Black), show a lack of awareness about them, and perpetuate their otherization. While we found instances of negative stereotypes against Whites (e.g., regarding their athletic ability), they occurred less frequently than stereotypes against racial minorities.
This asymmetry underscores the potential of search autocomplete feature to contribute to representational harm.

We found that both mainstream media and user-generated content contribute to the production of racial stereotypes in YouTube search outputs.
Notably, we observed that YouTube search 
% has several unique characteristics that promote group stereotypes. 
% For example, it 
amplifies distorted racial views that appear in mainstream entertainment media by reproducing titles or dialogues from existing media content (e.g., ``white men can't jump''). 
On the other hand, amateur YouTube creators, seemingly driven by the platform's commercial logic~\citep{boxman2019practice}, also contribute to racial stereotypes, often delivering them  
% Creators' communal and commercial aspirations~\citep{boxman2019practice} could also account for our finding that racial stereotypes on YouTube are often subtly delivered 
in the form of \textit{ethnic humor} and memes. 
Curiously, we found that such ethnic humor is sometimes delivered by creators belonging to racial minorities (e.g., Steven He O’Byrne), presumably to gain an audience by propagating stereotypes about their own race with impunity.

Besides racial stereotypes, we also identified traces of interracial tensions through various ways of cultural appropriation, anxieties around interracial relationships, and concerns about persistent underrepresentation in the media.
These insights reflect contemporary racial dynamics in the U.S. and suggest shifting terrains of racial discord, visibility, and exclusion that warrant further inquiry.

% Power dynamics
Balanced against the above problematic trends, we found that YouTube search outputs reproduce some creators' attempts to dismantle racial stereotypes, often through deploying humor (e.g., ``every black man should have a latte.''). 
However, such dismantling often occurred only after viewing the videos in the corresponding search results; in many cases, the autocomplete text alone---short, decontextualized, and hyper-visible---perpetuates racial biases (e.g., ``black woman can't buy a dress,'' ``white lives matter'').
We observed discursive events that showed a repudiation of racial inequalities and hopes for a post-race society (e.g., ``is there a hispanic disney princess''). 
We also found potential attempts at promoting positive representations of racial minorities (e.g., ``black people can dance to anything''),
seeking practical information about one's own race, and genuine curiosities about other races. Additionally, we found depictions of struggles and inequalities faced by racial minorities (e.g., ``asian men are bitter,'' ``can't keep a black man down'').
Thus, the user-generated content that drives YouTube serves to produce search outputs that go beyond racial stigmas and reveal the ideological messiness of the participatory culture~\citep{khan2022researching}.

% Ethical Deployment 1: nudge effect of biased autocomplete on users
Our results highlight ethical concerns about the potential `nudge effects' of autocomplete suggestions on YouTube users. 
\citet{graham2023ethical} noted that autocompletes can be biased in ways that might not be clear on an individual level but become apparent when evaluating a larger sample. Our analysis reveals evidence of similar \textit{aggregated discrimination} in race-related queries on YouTube.
For example, we find that while Black struggles to combat oppression are diminished, e.g., through the removal of references to ``Black Lives Matter'' in autocomplete suggestions, ``White Lives Matter'' continues to feature in these suggestions despite its known associations with hate groups.

Examining the relative prevalence of different themes in our data, we note that skin tone, which serves as an explicit marker of race, appeared much less frequently than other categories like culture and ability in search autocompletes. This aligns with prior findings by \citet{walters2024memes} on race-related Facebook memes, where explicit depictions of race and appeals to racism appeared much less frequently than implicit appeals to forward a racist agenda. This suggests a need to attend to traditionally palatable rhetorics of racism rather than just overtly racist speech on digital platforms. 

% Ethical Deployment 2: moderation strategy - platform how? 
Building on our findings, we recommend a careful development of moderation policies and practices for the ethical deployment of search autocomplete tools. 
% Prior research on moderation in search engine autocompletes has shown that platforms moderate derogatory stereotypes about various demographic categories \citep{leidinger2023stereotypes}. Google even has an autocompletion moderation policy to prevent autocompletes ``that might be violent, sexually-explicit, hateful, disparaging or dangerous'' \citep{sullivan2020}.  Yet, as our findings show, YouTube, owned by the same parent company as Google, perpetuates racial stereotypes, highlighting the need for improved policy and corporate-wide enforcement. 
As \citet{rogers2023algorithmic} points out, search platforms should take actions that exceed mere `quiet patches' of infamous autocompletions highlighted by journalists and researchers---they must institute overarching mechanisms that proactively detect and regulate problematic suggestions.
% ensuring that adequate remedies occur for all their products and language offerings. 
Like many AI-based technologies \citep{rai2020explainable}, autocomplete feature does not inform users how it generates specific outputs and why they embed identity-based biases.
Such lack of information allows search platforms to avoid taking the full ethical responsibility for problematic suggestions, even in cases when they result from machine learning biases rather than prior search queries or creators' contributions.
Therefore, we call for greater transparency regarding autocomplete sources and algorithms to facilitate a more precise attribution of problematic suggestions.

This study has some limitations.
We observed an under-representation of Hispanics in our analysis of autocompletes.
However, it is possible that this result is influenced by other terms referring to the ``Hispanic'' race (e.g., ``Latino,'' ``Mexican'') being used colloquially more often than similar counterparts (e.g., ``Caucasian,'' ``African-American'') for other races. 
Future research on autocomplete audits should also collect queries using such colloquial terms.
Our arguments about the influence of autocomplete biases on users would benefit from further engaging with the duality of how aggregated user behaviors and audience cues, in turn, feed into autocomplete outputs.
While we attempted to obtain a temporally stable and geographically representative dataset, the YouTube platform constantly updates its search autocomplete suggestions, thus limiting the study's generalizability.
Although outside the scope of the current study, future research would benefit from a systematic comparison of search autocompletes (and their associated biases) on Google and YouTube sites and examining the reasons behind those differences.

\section{Conclusion}
This article has presented the first systematic analysis of racial stereotypes in YouTube’s search autocomplete outputs, addressing a major gap given YouTube's cultural significance and its role as the second-largest search engine. Our thematic framework highlights asymmetric representation of different racial groups, evidence of harmful stereotypes, and hegemonic counter-struggles. 
We call for greater transparency and reform in content moderation to counteract identity-based harms enacted by search outputs.

% \subsection{Limitations and Future Work}
% This study has some limitations. First, YouTube autocomplete suggestions analyzed in this study might not be the same as the current status. This is because YouTube search predictions are affected by diverse factors like popularity that can change as time passes \citep{YouTubehelp}. Thus, some variations in autocomplete results for the same input queries are expected. Secondly, this study focused on the four racial groups to study racial stereotypes. We admit that these four groups do not represent the entire U.S. population. Future research can consider including more racial/ethnic groups. Finally, we focused on examining the biases toward race-based categorization. However, some autocomplete results partly reflect the impacts of intersectionality around diverse demographic groups including the combinations of race and gender. Since people are mutually affiliated with multiple identities such as gender and race, relying on a single identity may limit examining deeper and more comprehensive aspects of biases across demographic groups. Therefore, future research can consider taking an intersectional approach to scrutinizing more inclusive bias toward diverse social categories embedded in algorithmic systems.

%%%%%%%%%%%  Conclusions  %%%%%%%%%%%%%%

% \section{CONCLUSION}

% % %% The next two lines define the bibliography style to be used, and
% % %% the bibliography file.

\bibliographystyle{SageH}
\bibliography{bib.bib}

\clearpage
\setcounter{page}{1}
\setcounter{table}{0}
\setcounter{figure}{0}

\input{MainSupplement}

\end{document}

%% file: MainSupplement.tex
%%%%%%%%%%%  Supplemental Material  %%%%%%%%%%%%%%

% \newpage
\section{Examining Racial Stereotypes in YouTube Autocomplete Suggestions}
\begin{sm}

\noindent This document includes additional information on data preparation and descriptive statistics of the dataset. \\

\textbf{Data Preparation} \\
As part of our data preparation and preliminary analyses, we first conducted a comparative review of the collected autocomplete datasets. We analyzed data from all five US states collected three times to identify inter-location differences between autocomplete results. We used the data of New Jersey as a reference category because it had the highest number of autocomplete results compared to other states. Over 85\% of YouTube autocomplete results were consistent between New Jersey and the other four states: Texas (97.32\%), Georgia (95.54\%), California (91.67\%), and Washington (86.61\%). 

Given this high consistency, we determined to employ all the extracted data for data analysis after removing the duplicate autocomplete results from three different waves and five different states. 
% While this step precluded prioritizing more geographically and temporally stable autocompletes, it allowed us to work with a more comprehensive set of suggestions. 
This process of integrating datasets and removing the duplicates was carried out in three stages described below.

In the first stage, we integrated the data. 
We consolidated data extracted in three different waves for each state into one dataset per state, resulting in five datasets representing the five states. Then, the five datasets based on states were further merged back into one combined dataset (n=534). Accordingly, we obtained one unified dataset across three different waves and five different states. 
We tagged each autocomplete in our data with a racial category corresponding to the input query that produced that autocomplete.
Thus, the dataset comprised four racial categories. Figure \ref{fig:dataset} presents an example of autocomplete results for a few queries of the racial category `White' in our dataset. 

\begin{figure}[h]
   \centering
   \includegraphics[width=1\linewidth]{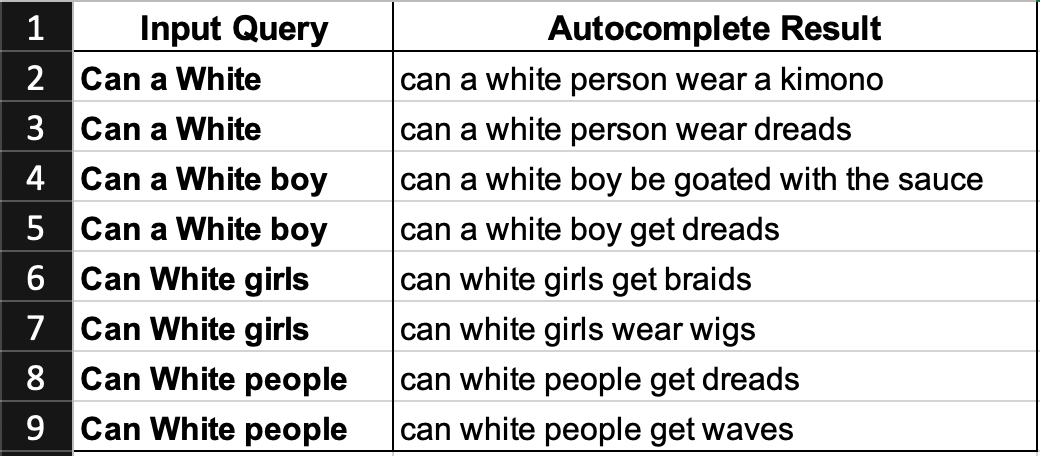}
   \caption{Examples of autocomplete results for the racial category ``White.''}
   \label{fig:dataset}
 \end{figure}

In the second stage, we processed our data to exclude the autocomplete suggestions that did not refer to race. For example, for the group terms relevant to ``White'' and ``Black'' races, several autocomplete results refer to the term as a color rather than a race. Thus, we used the following criteria used by \citet{baker2013white} to refine our dataset: whether the autocomplete suggestion refers to human subjects rather than non-human subjects. As a result, the dataset resulted in a total of 301 race-related autocomplete results.

In the last stage, we eliminated the autocomplete results that were produced multiple times. For example, the result ``how often should a black man wash his hair'' was generated from the input queries ``Should a Black,'' ``Should a Black man,'' and ``A Black man should.'' In such cases, following prior research~\citep{al2020google, baker2013white, roy2020age}, we removed duplicates and counted them as one result. Through this process, we obtained YouTube autocomplete results for the four racial groups without any duplicates. This led to a dataset of 241 unique autocomplete results on which we conducted our qualitative analysis. 

During this analysis, we further refined the dataset by removing autocompletes that belonged to the miscellaneous category, i.e., that were not included in our final two-tier category structure. As a result, our final dataset consisted of 217 autocomplete results. \\

\textbf{Descriptive Statistics} \\
%Final dataset
We calculated descriptive statistics for different datasets in our analysis. First, we calculated the number of generative input queries (i.e., queries returning at least one autocomplete suggestion related to racial groups) based on the dataset (n=301) that includes duplicate autocompletes (Table \ref{tab:new_tab}). 
We observed differences in the number of generative input queries across racial groups. 
% The number of generative input queries means the number of unique input queries that generate autocomplete related to racial groups. 
Whites and Blacks had the most generative input queries, followed by Asians and Hispanics. 
We also counted non-generative input queries, which refers to the queries that \textit{do not} generate any autocompletes regarding each racial group. We found large differences in the number of such input queries across racial groups. Surprisingly, 77 out of 84 (91.6\%) input queries did not generate any relevant autocompletes for Hispanics. Similarly, approximately 80\% of the input queries for Asians did not generate relevant autocompletes. Lastly, both White and Black categories had similar results---about 58\% of the input queries did not generate any autocompletes.

\begin{table}[ht]
  \centering
  \begin{tabular}{c|>{\centering\arraybackslash}p{3cm}|>{\centering\arraybackslash}p{3cm}}
    \toprule
    Category & \# of Generative input queries & \# of Non-generative input queries \\
    \midrule
    White & 36 & 48 \\
    Black & 35 & 49 \\
    Asian & 17 & 67 \\
    Hispanic & 7 & 77 \\
    \bottomrule
  \end{tabular}
    \vspace{0.3em}
  \caption{Descriptive statistics of generative and non-generative input queries.}
  \label{tab:new_tab}
\end{table}

% \begin{figure}
%     \centering
%     \includegraphics[width=1\linewidth]{no_auto_v2.png}
%     \caption{The number of autocompletes by race.}
%     \label{fig:auto_race}
% \end{figure}

% \begin{figure}[hb]
%     \centering
%     \includegraphics[width=1\linewidth]{no_input_queries_wo_auto_v3.png}
%     \caption{The number of input queries without any autocompletes by race.}
%     \label{fig:query_wo_auto_race}
% \end{figure}

Table \ref{tab:stats} presents the racial distribution in our dataset after the removal of duplicate autocompletes (n=241). Our result shows differences in the number of autocompletes across different racial groups --- 
`White' group had the highest number of autocompletes (n=99), followed by the `Black' group (n=95). This number dropped substantially for the Asian (n=37) and Hispanic (n=10) groups. 
This suggests that videos with Whites and Blacks are more visible, and the content including them is more likely to be recommended on this platform compared to Asians and Hispanics. This shows that YouTube under-represents minoritized racial groups in its autocomplete suggestions.

\begin{table}[ht]
  \centering
  \begin{tabular}{c|c}
    \toprule
    Category & \# of Autocompletes \\
    \midrule
        White & 99 \\
        Black & 95 \\
        Asian & 37 \\
        Hispanic & 10 \\
    \bottomrule
  \end{tabular}
    \vspace{0.3em}
    \caption{Racial distribution in our dataset after removal of duplicate autocompletes (n=241).}
    \label{tab:stats}
\end{table}

\newpage
Table \ref{tab:results} (on page 3) shows the racial distribution for the number of autocompletes in each subcategory following our qualitative analysis (n=217).
These results show that many categories disproportionately represent specific racial groups. 
For example, only Blacks have autocompletes that belong to \textit{Personal Hygiene}, \textit{Financial}, and \textit{Inappropriate Behavior} categories.

\begin{table*}[h]
  \centering
  \begin{tabular}{l l|c c c c}
    \toprule
    Main category & \ Subcategory & \ White \% (n) & Black \% (n) & Asian \% (n) & Hispanic \% (n) \\
    \midrule
        Appearance 
             % & Facial Hair & The hair on the face %(i.e., beard)  \\
             & Personal Hygiene 
             & - & 100 (6) & - & - \\
             & Skin Tone 
             & 10 (1) & 90 (9) & - & - \\
        Ability 
            & Talent
            & 72 (34) & 26 (12) & 2 (1) & - \\
            & Financial 
            & - & 100 (4) & - & -\\
            & Intellectual 
            & - & 75 (3) & 25 (1) & - \\
        Culture 
            & Cultural Heritage 
            & 78 (35) & 16 (7) & 7 (3) & - \\
        % & & (i.e., cultural practices and cultural attires) \\
        & Ethnic Humor 
            & 21 (6) & 18 (5) & 61 (17) & -  \\
        % & & within a specific region or cultural community  \\ 
        & Language 
            & 25 (3) & 42 (5) & 17 (2) & 17 (2)
        \\
        % & & (i.e., one’s voice or accent) \\ 
        % & Lifestyle & The typical pattern of individuals' or groups' lives \\
        % & & (i.e., daily behavior or habits)  \\
        & Relationships 
            & 47 (9) & 37 (7) & 11 (2) & 5 (1) \\
        % & & (i.e., romantic relationship) \\
        Social Equity 
            & Diversity/Inclusion 
            & 40 (2) & 40 (2) & - & 20 (1) \\
            & Racial Justice 
            & 36 (8) & 64 (14) & - & - \\
        Manner 
            & Aggression 
            & - & 82 (9) & 18 (2) & - \\
            & Inappropriate Behavior 
            & - & 100 (4) & - & - \\
    \bottomrule
  \end{tabular}
        \caption{The proportion of racial groups for each subcategory in our final dataset (n=217).}
        \label{tab:results}
\end{table*}

Note that YouTube’s autocomplete system does not strictly follow the exact syntactic patterns of input queries. Instead, it seems to generate suggestions based on broader associations between keywords. As a result, some of our collected data included autocomplete suggestions that do not include the original query patterns from which they were derived. Table \ref{tab:results2} (on the last page below) lists cases where this occurs and that we present as examples in our main manuscript.
This helps clarify the connections between our input queries and their corresponding outputs.

\begin{table*}[htpp]
  \centering
  \begin{tabular}{l l l r}
    \toprule
    Racial Group & \ Autocomplete Result & \ Input Query & \\
    \midrule
    White
        & how to do a white girl voice
             & Does a White girl \\
        & how to do a white person voice
             & Does a White person \\
        & do white lives matter one minute
             & Do Whites \\
        & do white lives matter shorts
             & Do Whites \\
        & that is a white man's luxury
             & Does a White man \\
        & can a white guy won [sic] an asian beauty pageant
             & Can a White \\
             & & Can a White boy \\
        & the white woman who turned black
             & A White woman is \\
        & white men can't jump 
             & White men can \\ 
        & white boys can't dance 
             & White boys can \\ 
        & white people can't dance 
             & White people can \\
    Black 
        & pretend there is a black man chasing you
             & Is a Black man \\
        & how to do a black person's hair
            & Does a Black person \\
        & how to do a black person's makeup
            & Does a Black person \\
        & how to do a black man bun
            & Does a Black man \\
        & how to do a black man voice
             & Does a Black man \\ 
        & how can a black person get curly hair
             & Does a Black person \\
             & & Can a Black person \\ 
             & & A Black person can \\
        & can't date a black woman until my grandpa died 
             & Does a Black woman \\
             & & A Black woman can \\
        & can't keep a black man down
             & Does a Black man \\
             & & Can a Black man \\ 
        & how often should a black man wash his hair
             & Should a Black \\
             & & Should a Black man \\
             & & A Black man should \\
        & how often should a black woman wash her hair
             & Should a Black \\
             & & Should a Black woman \\
             & & A Black woman should \\
        & how often should a black man wash his beard
             & Should a Black man \\
        & black woman can't buy a dress 
             & A Black woman can \\
        & a black man can't have a suitcase 
             & A Black man can \\
        & run like a black man is chasing you
             & A Black man is \\
        & every black man should have a latte
             & A Black man should \\
        & black man can't pronounce
             & Black men can \\
        & black man can't dance 
             & Black men can \\
        & black man can't play basketball 
             & Black men can \\
        & black man can't say beginning 
             & Black men can \\
        & black man cannot own g wagon 
             & Black men can \\
        & black man can't stop laughing
             & Black men can \\
        & black man cant [sic] hold his laugh
             & Black men can \\
        & black people can't marry meme
             & Black people can \\
        & black people can't marry white people
             & Black people can \\
        & black people can't name one african country
             & Black people can \\
        & black people can't hear smoke detectors
             & Black people can \\
    Asian 
        & what does an asian accent sound like
             & Does an Asian \\
        & what is an asian baby girl
             & Is an Asian \\
        & what is an abg asian baby
             & Is an Asian girl \\
        & when an asian is bad at maths
             & An Asian is \\
        & when an asian is kidnapped 
             & An Asian is \\
        & when an asian is president 
             & An Asian is \\
        & when an asian is your substitute 
             & An Asian is \\
        & asian man goes crazy 
             & An Asian man is \\
    Hispanic
        & is there a hispanic disney princess
             & Is a Hispanic \\
        & all latino accents 
             & A Hispanic should \\
        & woman with spanish accent 
             & A Hispanic girl should \\

    \bottomrule
  \end{tabular}
        \caption{This table lists autocomplete suggestions and the input queries they derived from for cases where the suggestion did not contain the input query. This highlights that YouTube’s autocomplete generation does not strictly follow the exact syntactic patterns of input queries.}
        \label{tab:results2}
\end{table*}

\end{sm}